\documentclass{article}
\usepackage{iclr2026_conference,times}
\usepackage{hyperref}
\usepackage{url}
\usepackage{amsmath,amssymb}
\usepackage{graphicx}
\usepackage{booktabs}
\usepackage{microtype}
\usepackage{xspace}
\usepackage{enumitem}
\usepackage{tikz}
\usetikzlibrary{arrows.meta,positioning,fit,shapes.misc,shapes.symbols}
\usepackage{tabularx}
\usepackage{array}     % for m{} column type
\usepackage{multirow} % in preamble
\usepackage{listings} % in preamble

\iclrfinalcopy

\makeatletter
\let\ICLRmaketitle\maketitle
\renewcommand{\maketitle}{%
  \ICLRmaketitle
  \fancyhead[L]{\textit{Preprint. This version is not peer reviewed.}}% overrides what the class set
}
\makeatother

\newcommand{\xmark}{--}

\newcommand{\model}{\textsc{BioVERSE}\xspace}
\newcommand{\biofm}{BioFM\xspace}
\newcommand{\biofms}{BioFMs\xspace}
\newcommand{\smallerllm}{Granite-8B\xspace}
\newcommand{\smallerllmfullname}{Granite-3.3-8B-Instruct\xspace}
\newcommand{\llm}{LLM\xspace}
\newcommand{\llms}{LLMs\xspace}

\newcommand{\BIO}{\texttt{[BIO]}}
\newcommand{\BIOi}[1]{\texttt{[BIO\_#1]}}
\newcommand{\trainbio}{\texttt{[TRAINABLE\_BIO]}}

\iclrfinalcopy

\title{\model: Representation Alignment of Biomedical Modalities to LLMs for Multi-Modal Reasoning}

\author{%
\small
Ching-Huei Tsou\textsuperscript{1}\thanks{Equal contribution. Corresponding author.} \quad
Michal Ozery-Flato\textsuperscript{2}\footnotemark[1] \quad
Ella Barkan\textsuperscript{2} \quad
Diwakar Mahajan\textsuperscript{1} \quad
Ben Shapira\textsuperscript{2} \\
{\small
\textsuperscript{1}IBM T.J. Watson Research Center, USA \quad
\textsuperscript{2}IBM Research, Haifa, Israel
} \\
\texttt{\fontsize{8}{10}\selectfont \{ctsou, dmahaja\}@us.ibm.com} \quad
\texttt{\fontsize{8}{10}\selectfont \{ozery, ella\}@il.ibm.com} \quad
\texttt{\fontsize{8}{10}\selectfont ben.shapira@ibm.com} \\
}
\begin{document}
\maketitle

\begin{abstract}
Recent advances in large language models (\llms) and biomedical foundation models (\biofms) have achieved strong results in biological text reasoning, molecular modeling, and single-cell analysis, yet they remain siloed in disjoint embedding spaces, limiting cross-modal reasoning. We present \model~(\textbf{Bio}medical \textbf{V}ector \textbf{E}mbedding \textbf{R}ealignment for \textbf{S}emantic \textbf{E}ngagement), a two-stage approach that adapts pretrained \biofms as modality encoders and aligns them with \llms through lightweight, modality-specific projection layers. The approach first aligns each modality to a shared \llm space through independently trained projections, allowing them to interoperate naturally, and then applies standard instruction tuning with multi-modal data to bring them together for downstream reasoning. By unifying raw biomedical data with knowledge embedded in \llms, the approach enables zero-shot annotation, cross-modal question answering, and interactive, explainable dialogue. Across tasks spanning cell-type annotation, molecular description, and protein function reasoning, compact \model configurations surpass larger \llm baselines while enabling richer, generative outputs than existing \biofms, establishing a foundation for principled multi-modal biomedical reasoning.
\end{abstract}

\section{Introduction}
High-throughput assays such as scRNA-seq, proteomics, and small-molecules profiling generate rich, high-dimensional data that are critical for biomedical discovery. Biomedical foundation models (\biofms; also referred to as BMFMs) trained on those inputs, e.g., scGPT \citep{scgpt2024} for single-cell RNA sequencing (scRNA-seq), ESM-2 \citep{esm2_2023} for proteins, Molformer \citep{molformer2022} for small molecules, capture expressive representations but lack instruction-following and open-ended reasoning. In contrast, general-purpose large language models (\llms) excel at language interaction and can nominally ingest sequences like proteins or Simplified Molecular Input Line Entry System (SMILES) strings, but tokenization yields short, uninformative fragments and they cannot parse modalities such as scRNA-seq, where a cell’s gene expression vector cannot be represented meaningfully as a token sequence. Bridging these strengths requires a framework that preserves modality-specific encoders, aligns their embeddings with the \llm token space, and enables reasoning across them.

We introduce \model, a framework adapting the familiar vision–language paradigm (e.g., Flamingo~\citep{flamingo2022}, BLIP-2~\citep{blip2-2023}, LLaVA~\citep{llava2023}), and more recently, InternVL3.5~\citep{internvl3-2025}, to the biomedical domain. \model follows a BioFM-adapter-LLM design: it projects \biofm embeddings into the \llm's embedding space via a lightweight MLP adapter and injects them as special tokens (e.g. \BIOi{1}, \BIOi{2}, ..., \BIOi{k}, and \trainbio). By placing biological and textual information in a shared space, \model enables joint multi-modal reasoning while directly exploiting the \llm's native memory and inference abilities. Our contributions are:
\begin{itemize}[leftmargin=*]
    \item \textbf{Modular architecture:} Plug-and-play biological encoders (scRNA-seq, protein, molecule) connect to a decoder-only \llm via a small projection layer and LoRA adapters.
    \item \textbf{Alignment via contrastive learning:} We directly align encoder embeddings to the language token space, i.e., no separate bio-language encoder, enabling zero-shot transfer across modalities.
    \item \textbf{Multimodal instruction tuning:} We curate paired (embedding, instruction, response) data so the \llm learns to use biological context in generation.
    \item \textbf{Practicality:} Compact \model variants match or exceed larger baselines on joint bio–text tasks and support privacy-preserving, on-prem deployments. This aligns with the current trend of promoting small language models in the agentic AI systems \citep{belcak2025small}.
\end{itemize}

% =========================================
% Related Work
% =========================================

\section{Related Work}
\paragraph{Biomedical encoders.} Large-scale \biofms have been developed for modality-specific data. For transcriptomics, models such as scBERT~\citep{scbert2022}, Geneformer~\citep{geneformer2023}, and scGPT~\citep{scgpt2024} capture cellular states and gene–gene dependencies, with BMFM-RNA~\citep{bmfmrna2025} providing a reproducible framework for pretraining. For proteins, ProteinBERT~\citep{proteinbert2022} and ESM-2~\citep{esm2_2023} learn contextual embeddings that support function and family prediction, while AlphaFold~\citep{alphafold2021} and ESMFold~\citep{esm2_2023} show how such embeddings enable structure prediction. In the molecular domain, ChemBERTa~\citep{chemberta2020} and MolFormer~\citep{molformer2022} encode SMILES strings and molecular graphs into embeddings of chemical properties. Together, these unimodal encoders yield strong representations but lack natural-language reasoning.

\paragraph{Bio-LLM integration.} Several efforts have explored bridging biological embeddings with language models; however, current methods only partially address the challenge of joint bio-text reasoning. GenePT~\citep{genept2024} pools gene-level embeddings derived from ChatGPT descriptions into cell-level representations, which work well for classification but are not integrated into an \llm’s generation pipeline.
CELLama~\citep{cellama2024} prompts LLMs with transcriptomic profiles converted into text-like inputs. While effective for flexible queries, it does not exploit pretrained \biofms trained directly on raw scientific data, limiting its ability to capture domain-specific signal.
scCello~\citep{sccello2024} and scMulan~\citep{scmulan2024} incorporate text labels or metadata as supervision to improve biological embeddings, but the resulting embeddings remain modality-specific and are not aligned with text embeddings from \llms, limiting their use for joint bio–text reasoning.
CellWhisperer~\citep{cellwhisperer2024} uses Geneformer~\citep{geneformer2023} and BioBERT~\citep{biobert2020} to align scRNA-seq and text embeddings. While this enables retrieval, differences in tokenization and architecture prevent seamless integration with generative LLMs, leading to a RAG-style pipeline rather than embedding-aware reasoning.
TxGemma~\citep{txgemma2025}, BioT5~\citep{biot52023}, and Galactica~\citep{galactica2022} fine-tune general-purpose or biomedical LLMs on biological sequences tokenized as text (e.g., amino acids, SMILES, or curated biomedical corpora). This design enables strong domain-specific reasoning and therapeutic applications but constrains the models to operate entirely in the text-token space. As a result, they do not leverage pretrained BioFMs trained on raw molecular or cellular data, limiting their ability to capture low-level biological signals and reducing extensibility across modalities.
MAMMAL~\citep{mammal2024} unifies multiple bio-modalities and supports generation in a T5-style foundation model trained end-to-end on diverse data, but its monolithic nature and custom tokenizer preclude modular embedding reuse or deployment within modern instruction-tuned \llms.

\paragraph{General multi-modal LLMs.} 
In the general AI domain, vision-language models demonstrate how non-text modalities can be modularly aligned with LLMs. Approaches like LLaVA, BLIP-2, Flamingo, and more recently InternVL 3.5 \citep{llava2023, blip2-2023, flamingo2022, internvl3-2025} established a design pattern where modality-specific encoders are efficiently connected to LLMs through projection and instruction tuning. This pattern has yet to be fully realized in biomedicine, where biological and text embeddings remain misaligned.

\paragraph{Positioning of \model.} Building on the proven encoder-projector-LLM design pattern from vision-language models, \model addresses the gap between biological and textual embedding spaces by projecting \biofm outputs directly into the \llm's input embedding space. This modular alignment enables pretrained encoders for scRNA-seq, proteins, or molecules to be integrated without retraining the \llm. By treating these embeddings as first-class tokens, \model allows the model to reason jointly over biological data and natural language, providing a flexible foundation for cross-modal biomedical intelligence.

\section{Method}
% ---------- Figures (placeholders) ----------
\begin{figure}[t]
    \centering
    % Replace with actual path in Overleaf project
    \includegraphics[width=0.9\linewidth]{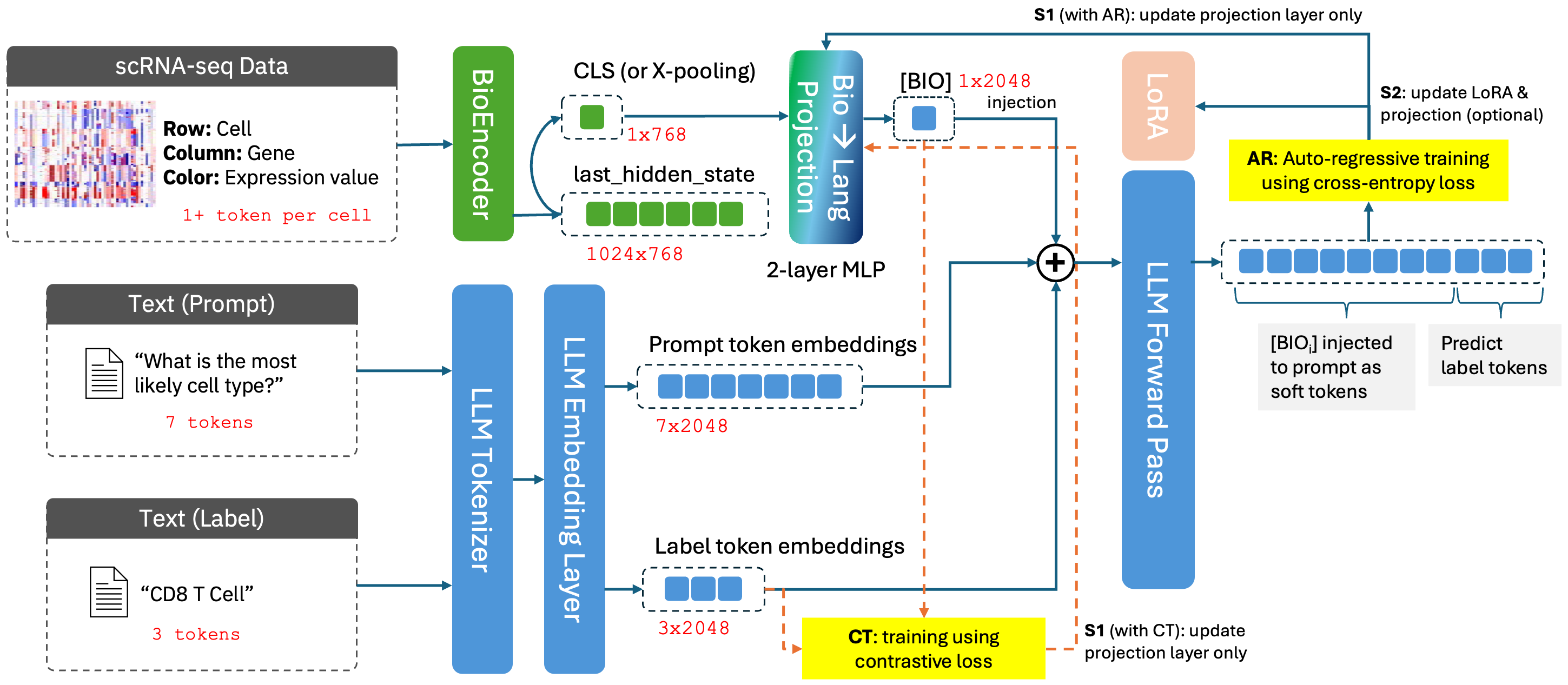}
    \caption{\model base architecture: a modality-specific \biofm encodes a biological entity, and its output embeddings are mapped by a projection layer into the \llm's embedding space via special tokens (e.g. \BIO). In the alignment stage, only the projection layer \(P_\theta\) is trainable, while the encoder \(f_b\) and the \llm \(g\) remain frozen. In the subsequent instruction-tuning stage, we allow both \(P_\theta\) and the low-rank adapter (\textsc{LoRA}) within the \llm to be trainable. Stage 1 (S1) can be trained using autoregressive (AR) or contrastive (CT) loss, while stage 2 (S2) is always AR.}
    \label{fig:arch}
\end{figure}

\subsection{Problem Setup}
Given a biological input \(x_b\) (e.g., a protein sequence or an scRNA-seq profile) and a natural-language context or query \(q\), our goal is to enable a frozen \llm to jointly reason over \((x_b, q)\).
We use a pretrained \biofm \(f_b\) to encode \(x_b\) into one or more embeddings \(z_b\), and a lightweight projection \(P_\theta\) to map \(z_b\) to the \llm's token-embedding space. These projected embeddings are injected at designated marker positions, e.g., \BIO, and act as soft tokens that the frozen decoder can attend to. The key challenge is that bio embeddings and text embeddings are trained in siloed spaces and must be aligned for effective joint reasoning.

\subsection{Two-stage Training}
\paragraph{Alignment}
Although \llm can, in principle, learn cross-space attention through sufficient instruction tuning, pre-aligning the bio and language embeddings provides a strong inductive prior: it reduces task-specific tuning burden and improves zero/few-shot generalization to unseen tasks. 
To achieve this, we introduce a CLIP-style alignment stage using paired data \((x_b, t_b)\), where the projection \(P_\theta\) is trained so that the bio embedding \(z_b = f_b(x_b)\) is close to its language counterpart \(\phi(t_b)\), which represents the text's embedding.

In the base training mode illustrated in 
Figure~\ref{fig:arch}, all data is processed through \llm's forward pass, and a standard autoregressive cross-entropy loss (with teacher forcing) is used to guide learning.
Alternatively, to avoid the costly forward pass required by a large \llm{}, and to enable alignment with efficient encoder models that are often co-trained with the decoder for retrieval, we evaluate an alternative alignment mode. In this variant, we use contrastive learning to directly align the bio embeddings with their paired text embeddings. From here onward, we denote the first alignment strategy as AR (autoregressive) and the second as CT (contrastive).
% The key variation lies in how \(\phi(t_b)\) is constructed, whether from static token embeddings, paired encoders, or decoder states. We therefore study two axes: (i) alignment strategies between \biofm and \llm, and (ii) the choice of \(\phi(.)\) as the anchor for alignment.
\paragraph{Instruction Tuning}
The alignment stage maps \biofm embeddings into the \llm's token space; the instruction tuning stage teaches the decoder to use those soft tokens under real prompts, improving generative reasoning, prompt robustness, and likelihood calibration. Abundant curated corpora (e.g., TxGemma-processed instruction sets and Therapeutics Data Commons (TDC) tasks) can be readily adapted for multi-task SFT, making this step practical with minimal data engineering. However, instruction tuning can confound alignment comparisons, increase compute and hyperparameter burden, and cause geometry drift. Given that our focus is alignment, we do not perform extensive instruction tuning.

\subsection{Modular Architecture}
\paragraph{Biological Encoder}
We use a pretrained \biofm \(f_b: \mathcal{X}_b \!\to\! \mathbb{R}^{k\times d_b}\) to encode a biological input \(x_b\) into \(k\) embeddings:\footnote{For notational simplicity, we assume $k=1$ and omit the token index in most equations from here onward; the framework naturally extends to $k>1$ when multiple embeddings are injected}
\[
z_b = f_b(x_b), \quad z_b \in \mathbb{R}^{k \times d_b}
\]
Any \biofm appropriate to the modality can be used, provided a bio embedding can be extracted from its output or an intermediate layer. We refer to it as an encoder to highlight its role in mapping a biological entity into an embedding, although the underlying architecture may be an encoder or a decoder. 
Some \biofms return a single pooled embedding (e.g., scGPT \citep{scgpt2024}, scBERT \citep{scbert2022}), while others output a sequence of contextual embeddings (e.g., per-residue embeddings in ESM-2 \citep{esm2_2023}, or per-token embeddings over SMILES in ChemBERTa \citep{chemberta2020}. In practice, these sequence outputs are often pooled to obtain a single vector per entity, but our framework supports both pooled and multi-token cases.
 
\paragraph{Projection Layer}
The projection $P_\theta : \mathbb{R}^{d_b} \to \mathbb{R}^{d_t}$ is a lightweight MLP with ReLU activations, layer normalization, and dropout for stability. It maps the \biofm output into the \llm embedding space:
\[ 
\tilde{z_b} = P_\theta(z_b), \quad \tilde{z_b} \in \mathbb{R}^{k \times d_t}
\]
% A simple residual pathway is added to improve robustness when dimensions match. 
In vision-language models, tens to hundreds of tokens are used per image (e.g., BLIP-2 \citep{blip2-2023}, CLIP \citep{clip2021}), often requiring token-level normalization or gating for stability. By contrast, \biofms usually pool to a single token, reflecting that cells, proteins, and molecules are typically treated as indivisible units, and a lightweight projection is sufficient to ensure compatibility with the \llm while preserving \biofm semantics. 
\paragraph{Injection of Bio Tokens}
We inject the projected bio embeddings $\tilde{z_b}$ at a placeholder (e.g., \BIO) within the query $q$, replacing the marker with the embeddings as soft tokens before concatenating with the rest of the sequence:
\[
[\, \text{Tokens}(q, \BIO \rightarrow \tilde{z_b}) \; ; \; \text{Tokens}(t_b) \,].
\]
Here, $\text{Tokens}(\cdot)$ denotes text after tokenization and embedding lookup. While standard text inputs are mapped from token IDs through the embedding matrix, projected bio embeddings are fed directly into the \llm embedding layer (via the \texttt{inputs\_embeds} interface in many implementations), enabling integration without modifying the tokenizer or embedding matrix. 

\paragraph{Language Embedding Targets}
When performing alignment training using AR loss, the model consumes $\text{Tokens}(t_b)$ directly, since 
the objective is next-token prediction over a text sequence. However, to also support CT loss that enforces representation-level similarity, we require a single pooled representation of the text. We therefore define a frozen language embedding \(\phi(t_b) \in \mathbb{R}^{d_t}\) extracted from the target \llm. Choices of the target will be discussed later in detail.

\paragraph{Language Model}
We extend a small \llm \(g\) with a few soft tokens, without modifying its tokenizer or positional encodings. As the lightweight projection layer decoupled the \llm and requires only embedding dimension compatibility, our approach can directly scalable to larger \llms.

\subsection{Alignment Objectives}
\paragraph{Autoregressive Decodability.}
Our default alignment strategy is to directly train the LLM to use the projected bio embeddings during generation. 
Given a query $q$, bio embeddings $\tilde{z_b}$ injected at a \BIO marker, and paired target text $t_b = (t_1, \dots, t_{|t_b|})$, we minimize the negative log-likelihood of predicting $t$ in an autoregressive manner:
\[
\mathcal{L}_{\mathrm{AR}}
= - \sum_{i=1}^{|t_b|} \log p_\mathrm{LLM}\big(t_i \mid \tilde{z}_b, q, t_{<i}\big)
\]
This objective explicitly teaches the LLM to attend to bio tokens in the same way it attends to text tokens, ensuring decodability and downstream reasoning ability. 
Because the loss is defined over natural text generation, it tightly couples alignment with the LLM’s causal decoding process.

\paragraph{Contrastive Alignment.}
In addition to autoregressive decodability, we also study a contrastive alignment mode that enforces representation-level similarity.
Here, the projected bio embeddings $\tilde{z_b}$ are aligned with text embeddings $\phi(t_b)$ from paired descriptions using a bidirectional InfoNCE loss:
\[
\mathcal{L}_{\text{CT}} = -\frac{1}{2N} \sum_{i=1}^N 
\Bigg[
\underbrace{\log \frac{\exp(\text{sim}(\tilde{z_b}^{(i)}, \phi(t_b^{(i)}))/\tau)}
{\sum_{j=1}^N \exp(\text{sim}(\tilde{z_b}^{(i)}, \phi(t_b^{(j)}))/\tau)}}_{\text{bio}\to\text{text}}
+
\underbrace{\log \frac{\exp(\text{sim}(\phi(t_b^{(i)}), \tilde{z_b}^{(i)})/\tau)}
{\sum_{j=1}^N \exp(\text{sim}(\phi(t_b^{(i)}), \tilde{z_b}^{(j)})/\tau)}}_{\text{text}\to\text{bio}}
\Bigg]
\]
where $\text{sim}(\cdot,\cdot)$ denotes cosine similarity and $\tau$ is a learnable temperature. 
Note that the denominator from bio to text normalizes over all text embeddings, and the denominator from text to bio normalizes over all bio embeddings. Prior work (e.g. CLIP \citep{clip2021}, BLIP-2 \citep{blip2-2023}) shows including both directions stabilizes training.

We explore contrastive alignment for three main reasons: 
(1) it enforces semantic consistency between bio and text embeddings rather than relying solely on next-token prediction, which may improve generalization to unseen tasks; 
(2) it decouples alignment from the frozen LLM decoder, allowing alternative text encoders to serve as alignment targets; 
and (3) it is computationally efficient, bypassing the \llm's forward pass to directly align paired $(x_b, t_b)$ examples. 
An additional benefit, observed in prior work, is that contrastive objectives produce more isotropic embedding spaces and can exploit large in-batch negatives, improving transfer and data efficiency.

\section{Experimental Setup}

\subsection{Models}

\paragraph{Biological Encoder}
We evaluate representative foundation models across three modalities (all pooled into a single embedding at the end): scGPT \citep{scgpt2024} for scRNA-seq, ESM-2 \citep{esm2_2023} for proteins, and ChemBERTa \citep{chemberta2020} for small molecules. We also include MAMMAL \citep{mammal2024}, a multimodal biomedical model that supports all three modalities. Finally, we consider a general-domain \llm used directly as a bio encoder. Although \llms can ingest serialized versions of biological entities (e.g., scRNA-seq approximated by sorting genes into a sequence, while proteins and SMILES strings are natively sequential), the resulting tokenizations tend to be short and poorly contextualized, and we suspect it limits biological fidelity.
\paragraph{Language Model}
We evaluate two scales of LLMs as the language backbone. As a small model, we use \smallerllmfullname (\smallerllm for short), an 8B open-weights model released by IBM Research, to demonstrate alignment effectiveness under limited capacity. \model is \llm-agnostic: any model that accepts embedding inputs (as is the case for most HuggingFace \llms) can be used without architectural changes.\footnote{In practice, this requires only that the LLM expose an \texttt{inputs\_embeds} interface or equivalent.} For comparison against large-scale baselines that do not leverage \biofms for encoding, we also evaluate GPT-OSS-120B, a public 120B open-weights model by OpenAI, details of models see Appendix.

\paragraph{Projection Layer}
The projection is implemented as a three-layer MLP with ReLU activations, layer normalization, and dropout for stability. 

\paragraph{Language Embedding Target}
We evaluate four ways to construct \(\phi(t)\): (1) \textbf{TokEmbed}, averaging input embeddings; (2) \textbf{LL-Mean}, mean pooling of the final layer; (3) \textbf{LayerAvg}, averaging several top layers; and (4) \textbf{LLM-Embed}, a co-trained text encoder when available. TokEmbed performed poorly, and we adopt LL-Mean as the default. LLM-Embed shows strong results but applies only under contrastive training; systematic study of LayerAvg and LLM-Embed is left for future work.

\subsection{Datasets and Evaluation}
\subsubsection{Alignment}
\label{sec:Alignment}
For alignment, we construct paired biological entities $(x_b)$ and textual descriptions $(t_b)$ across three modalities.
\textbf{Protein:} We obtain protein–text pairs from UniProtKB, where each amino acid sequence is linked to curated Gene Ontology (GO) terms representing its functional annotations across the three GO namespaces: Biological Process, Molecular Function, and Cellular Component. GO term metadata is derived from the official GO ontology, and annotations are obtained from UniProt cross-references. To ensure reliability, we retain only experimentally supported GO annotations, yielding high-quality supervision for aligning BioFM protein embeddings with language representations.
\textbf{Small Molecule:} For small molecules, we leverage LLASmol \citep{llasmol-yu2024}, which provides SMILES–text pairs with chemically grounded descriptions. Specifically, we select two datasets from the LLASmol collection for BioFM alignment: SMILES-to-IUPAC conversion and molecule captioning. In both cases, each molecule is represented as a SMILES string paired with natural-language annotations of structure, properties, or activities.
\textbf{scRNA-seq:} For single-cell data, we adopt CellWhisperer \citep{cellwhisperer2024}, which aligns scRNA-seq profiles with cell-type and tissue-level textual metadata. Following the dataset protocol, we use the CellxGene subset \citep{perkel2024cellxgene}, where pseudo-bulk RNA samples are generated by averaging single-cell profiles, and natural-language descriptions are produced from cell and tissue metadata using large language models. This enables alignment between transcriptomic embeddings and ontological descriptions.

\subsubsection{Instruction Tuning}
For Stage~2 instruction tuning, we augment the alignment dataset with templated prompts paired to the biological–text examples. This teaches the \llm to use aligned bio tokens under user queries; for example,
\quad \textit{"What cell type matches this \BIO gene-expression profile?"}
Since the primary goal of this paper is to evaluate architectural design rather than extensive prompt handling, we limit instruction tuning to light augmentation. This procedure can be readily extended with training data from prior works such as TxGemma \citep{txgemma2025}, which introduces instruction-style data for therapeutic reasoning tasks using the Therapeutics Data Commons (TDC), and CellWhisperer \citep{cellwhisperer2024} for cell-related tasks.

\subsection{Evaluation}
We evaluate our approach on six downstream tasks: five from Mol-Instruct \citep{mol-instruct-fang2023} (four protein-related and one small-molecule) and one from scEval \citep{sceval-liu2023} (cell-type annotation). Mol-Instruct provides molecular question–answer pairs spanning property prediction, reaction reasoning, and therapeutic relevance, while scEval offers benchmarks for scRNA-seq applications.
For generative tasks, we report results using three complementary metrics. \textbf{LLM-as-a-judge:} GPT-OSS-120B scores each model response independently against the expected output with single-output, reference-based prompt (see Appendix, repeating each evaluation three times under different random seeds. \textbf{BERTScore:} captures semantic similarity. \textbf{ROUGE-L:} measures surface-form overlap. Full definitions of the metrics and the prompt used for LLM-as-a-judge are provided in Appendix. Training details (learning rate, batch size, optimizer, training duration, and compute resources) are documented in Appendix.

\section{Results}
\subsection{Embedding Alignment Visualization}
\begin{figure}[t]
\centering
\includegraphics[width=0.46\linewidth]{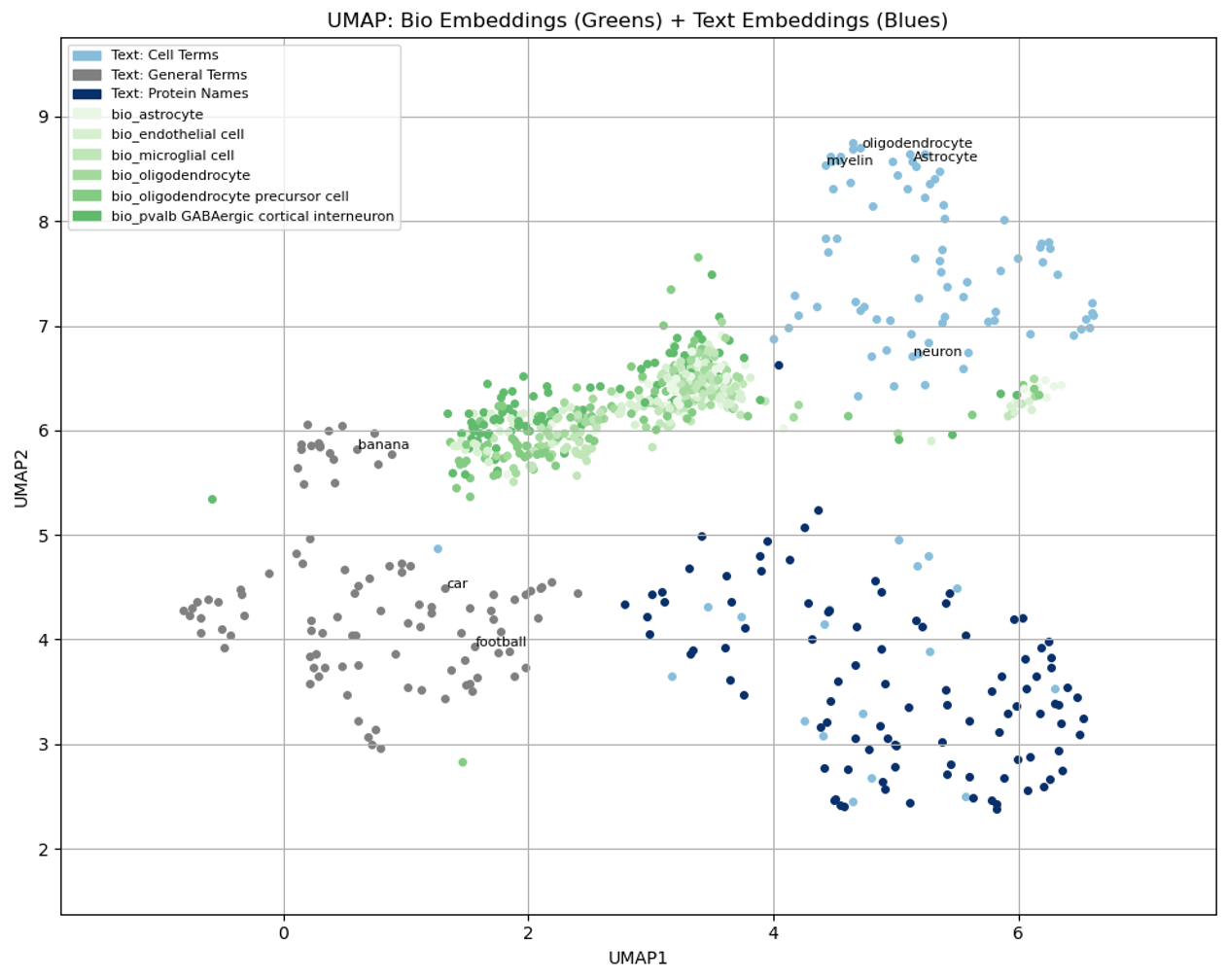}
\hfill
\includegraphics[width=0.46\linewidth]{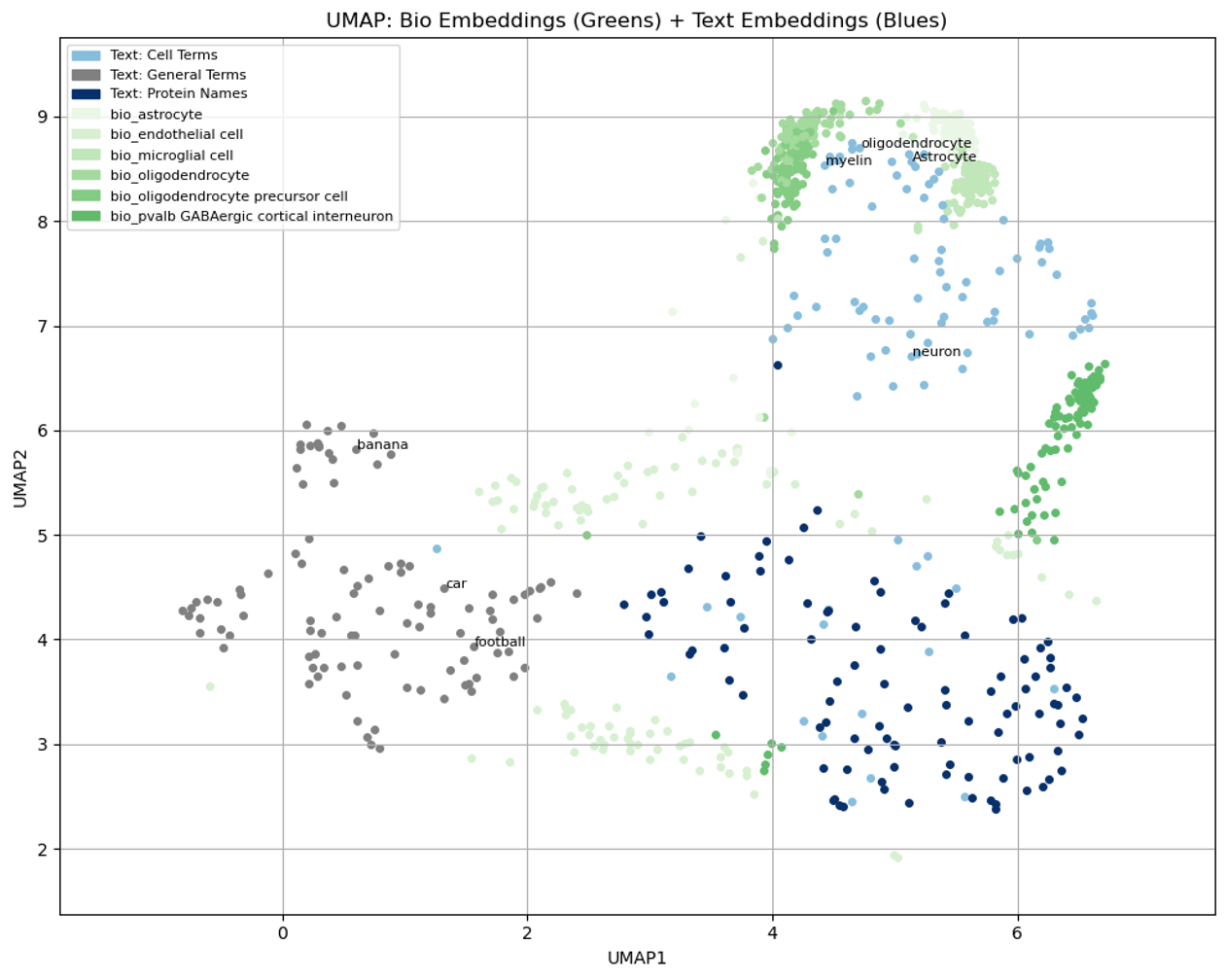}
\caption{UMAP visualization of scRNA-seq and text embeddings. Left: before alignment, cell embeddings (green) form isolated clusters within the \llm embedding space. Right: after alignment, cell embeddings are pulled closer to biologically relevant text and separated from unrelated general-domain text. \model successfully realigns the modalities into a shared representation space.}
\label{fig:umap_alignment}
\end{figure}

To illustrate the effect of our alignment procedure, we present UMAP projections of scRNA-seq embeddings and their corresponding natural language embeddings before and after applying BioVERSE alignment. As shown in Figure \ref{fig:umap_alignment}, prior to alignment, the two modalities occupy largely disjoint regions of the latent space, whereas after training the projection layer they exhibit clear overlap, indicating successful cross-modal alignment. These visualizations serve as a qualitative preview of BioVERSE’s capacity to unify biological and textual representations.

% ---- Figure placeholder for UMAP before/after ----

\subsection{Main Results}
\subsubsection{Zero-shot Generative Cell Type Annotation}

We evaluate \model's against two baselines on the PBMC10K dataset as discussed in scvi-tools \citep{scvi-gayoso2022python} under zero-shot generation: (1) random and majority baseline (2) open-domain \llms given a list of the 128 most expressed genes (sorted by expression count) as input.
Alignment is trained on CellxGene data aggregated into pseudo-bulk samples as in CellWhisperer \citep{cellwhisperer2024}. The PBMC10K dataset used for evaluation is not present in CellxGene; however, CellxGene contains another PBMC dataset among its 1,800+ scRNA-seq datasets. Thus, while the exact test set is excluded, the ontology of cell types is shared. This reflects a realistic zero-shot transfer setting. 
\begin{table}[t]
\centering
\small
\caption{Zero-shot PBMC10K results with 9 cell types.}
\label{tab:scrna_results}
\begin{tabular}{l|cc|c|ccc}
\toprule
& \multicolumn{2}{c|}{\textbf{Baseline}} & \textbf{Matching} & \multicolumn{3}{c}{\textbf{Generative}} \\
& Random & Majority & LangCell & \smallerllm & GPT-OSS-120B & \model \\
\midrule
Accuracy & 0.111 & 0.417 & 0.865 & 0.369 & 0.779 & 0.614 \\
Macro $F_1$ & 0.086 & 0.065 & 0.896 & 0.262 & 0.543 & 0.437 \\
\bottomrule
\end{tabular}
\end{table}

As shown in Table~\ref{tab:scrna_results}, although majority voting achieves relatively high accuracy, it fails on minority classes, leading to poor macro-$F_1$. 
Prior \biofms such as scGPT \citep{scgpt2024} and alignment-based models like LangCell \citep{zhao2024-langcell} and scMMGPT \citep{scmmgpt-shi2025language}, when performing cell type annotation under zero-shot setting, fundamentally operate in a candidate-space matching paradigm. These models project cells and a predefined set of candidate labels and their descriptions into a shared embedding space and assign the nearest match. LangCell achieves the highest scores, reflecting the relative ease of candidate-space matching. 
By contrast, generative models operates in a generative regime: the \llm must produce a natural language label rather than selecting the nearest candidate. In our setup, we apply prompt-level constraints, instructing the model to select only from a predefined option set without decoding-level enforcement. The model nevertheless engages in open-ended reasoning before aligning to a candidate, making the task inherently more difficult. This setting, however, offers unique advantages: the ability to articulate rationales, propose novel labels outside a fixed ontology, and integrate bio-embeddings with broader biomedical knowledge.

Open-domain \llms perform substantially better than chance, indicating that even with only sorted gene lists, \llms show some inherent capability for this task. \model improves substantially over its backbone (\smallerllm) while preserving and enhancing the \llm's reasoning ability when both \BIO and gene-list evidence are provided in the prompt. The \BIO token guides the model toward the correct type, but crucially also anchors the explanation to biological features, yielding more faithful rationales than using gene lists alone. While overall accuracy still trails candidate-matching approaches, the generative setting enables richer outputs: models can articulate why a label was chosen, highlight relevant genes, and remain extensible to novel types outside a fixed ontology. Future work will explore strengthening this interpretive capacity (e.g., through multiple \BIO tokens tied to specific pathways or gene modules) and scaling aligned projections to larger \llms.

\begin{figure}[t]
\centering
\begin{minipage}{0.96\linewidth}
\begin{lstlisting}[basicstyle=\scriptsize\ttfamily, frame=single]
True Label: CD14+ Monocytes
Predicted Label: Based on the sorted expressed genes, the most likely immune cell subtype 
is CD14+ Monocytes. The presence of genes such as TYROBP (DAP12), FCER1G (FcgRI), ITGB2 
(CD29), and ITGAM (CD11b) suggests a monocytic lineage. [...skip] The absence of B cell-
specific genes and T cell receptor genes (TR genes) further supports this conclusion.
\end{lstlisting}
\end{minipage}
\caption{Example generative annotation on PBMC10K: \model produces the label and reasoning grounded in gene evidence.}
\label{fig:gen_annotation_example}
\end{figure}

% -------------------------------------------------
% -------------------------------------------------
\subsubsection{Molecule Description Generation}
The molecular description generation task in Mol-Instructions \citep{mol-instruct-fang2023} evaluates a model’s ability to produce detailed free-text descriptions of molecules given their SMILES representation. Target outputs cover structural features, physicochemical properties, biological activities, and potential applications, requiring the model to bridge symbolic chemical notation with natural language.  
We compare \model with open-weight LLMs ranging from 8B (the same size as the \model backbone) to 120B, all without \biofm alignment and therefore relying only on raw tokenized SMILES strings.
We also test on the effect of using different \biofms (ChemBERTa vs.\ MAMMAL) to generate the initial molecular embeddings. This tests whether \model can flexibly adapt to modality-specific encoders without losing stability.  
All evaluations are conducted in a \emph{zero-shot transfer} setting: Mol-Instructions descriptions are not used during alignment. Instead, \model is aligned on independent molecule–text pairs from LLASMol \citep{llasmol-yu2024}, as described in Section~\ref{sec:Alignment}.  
Table~\ref{tab:mol_results} shows \model outperform open-domain \llms significantly, regardless of the size. Switching from MAMMAL to ChemBERTa yields slightly worse results under the same training iterations, indicating that the framework is plug-and-play and stable across different molecular encoders. Additionally, the two-stage strategy (S1 followed by S2) is more effective than simply training S1 for longer. 
%================================================
%================================================
\begin{table}[t]
\centering
\small
\caption{Molecular description generation results. S1: projection-only. S2: projection+LoRA}
\label{tab:mol_results}
\begin{tabular}{l c c c c c c}
\toprule
Model & BioFM & S1 & S2 & LLM-J & BERT-S & ROUGE-L \\
\midrule
\multirow{2}{*}{\model} & \multirow{2}{*}{MAMMAL}  & 30k      & 30k & \textbf{0.17}  & \textbf{0.92}  & \textbf{0.20} \\
& & 30k      & \xmark     & 0.10  & 0.92     & 0.18 \\
\hline
\multirow{2}{*}{\model} & \multirow{2}{*}{ChemBERTa} & 130k & \xmark & 0.10 & 0.91 & 0.18 \\
&  & 30k  & \xmark & 0.08 & 0.90 & 0.16 \\
\hline
\smallerllm  & \multicolumn{3}{c}{\multirow{4}{*}{(not applicable)}} & 0.04 & 0.91 & 0.07 \\
LLaMA-70B    & \multicolumn{3}{c}{}                        & 0.05 & 0.90 & 0.06 \\
Mixtral-8x7B & \multicolumn{3}{c}{}                        & 0.05 & 0.91 & 0.08 \\
GPT-OSS-120B & \multicolumn{3}{c}{}                        & 0.02 & 0.89 & 0.06 \\
\bottomrule
\end{tabular}
\end{table}
All three evaluation metrics show a consistent trend across our tests. We consider LLM-J to be the most meaningful metric for free-text generation; we therefore report only this metric in subsequent results. 
% -------------------------------------------------
% -------------------------------------------------
\subsubsection{Protein-oriented Text Generation}
We evaluate all four protein-oriented text generation benchmarks from Mol-Instructions \citep{mol-instruct-fang2023}: 
(1) catalytic activity prediction, 
(2) domain/motif prediction, 
(3) functional description generation, and 
(4) protein function prediction. 
Each task provides a protein sequence as input, and the model must generate free-text outputs describing a specific property of that sequence. Together, these tasks probe both factual grounding (e.g., motif recognition) and open-ended description ability, testing whether the model can jointly reason over the protein sequence and the accompanying prompt.

Similar to the molecular task, we compare \model with open-weight LLMs without \biofm alignment and therefore relying only on raw tokenized amino-acid sequences. We also conduct self-comparisons along two axes: (1) training iterations and (2) alignment strategies. 
As shown in Table \ref{tab:protein_results}, across all four tasks, \model consistently outperforms open-domain LLMs by a wide margin. Longer alignment training further improves results, and the two-stage strategy, i.e., first training the projection (S1), then training projection and LoRA jointly (S2), yields the strongest performance. For instance, (30K S1 + 30K S2) outperforms (100K S1), and (100K S1 + 100K S2) outperforms (500K S1) in one task and achieved comparable overall scores in our benchmarks. Switching from MAMMAL (458M parameters) to a small ESM2 (8M parameters), the performance dropped, highlighting the impact of the encoder's quality. When CT is used to reduce the training time in S1, it is important to follow it with S2, as S1-only does not teach the \llm backbone how to use those tokens in a generative task, and when combined with a prompt results in unexpected generation. However, a small S2 quickly bring up the performance of CT and with 30K S2 the performance is comparable to the longest run with AR.  
All results are reported in a zero-shot transfer setting. \model is aligned (both S1 and S2) only on UniProtKB protein-text pairs with short GO terms and curated annotations, while evaluation is performed on the Mol-Instructions test split, which requires long-form, free-text property descriptions. This ensures that performance reflects transfer beyond the ontology terms used during alignment. 
% ---- Table 2 ----
\begin{table}[t]
\centering
\small
\caption{Protein text generation tasks results. All scores are LLM-J}
\label{tab:protein_results}
\setlength{\tabcolsep}{6pt}
\begin{tabular}{l r r r r r r r r r}
\toprule
\textbf{Model} & \textbf{\biofm} & \textbf{Align.} & \textbf{S1} & \textbf{S2} & \textbf{catal.} & \textbf{motif} & \textbf{func.} & \textbf{prot.} & \textbf{Avg.} \\
\midrule
\multirow{3}{*}{\model} & \multirow{3}{*}{MAMMAL} & \multirow{3}{*}{AR} &  500k & 500k & \textbf{0.37} & \textbf{0.21} & \textbf{0.40} & 0.35 & \textbf{0.33} \\
& & & 100k & 100k & 0.35 & 0.19 & 0.38 & 0.32 & 0.31 \\
& & & 30k &   30k & 0.32 & 0.18 & 0.33 & 0.29 & 0.28 \\
\hline
\multirow{3}{*}{\model} & \multirow{3}{*}{MAMMAL} & \multirow{3}{*}{AR} & 500k &  --  & 0.34 & 0.20 & 0.38 & 0.38 & 0.32 \\
& & & 100k &  --  & 0.26 & 0.17 & 0.33 & 0.32 & 0.27 \\
& & &  30k &  --  & 0.21 & 0.11 & 0.22 & 0.31 & 0.21 \\
\hline
% \model & MAMMAL & CT & 500k &  500k  & -- & -- & -- & -- & -- \\
% \model & MAMMAL & CT & 100k &  100k  & -- & -- & -- & -- & -- \\
\multirow{2}{*}{\model} & \multirow{2}{*}{MAMMAL} & \multirow{2}{*}{CT} & 30k &  30k  & 0.33 & \textbf{0.21} & 0.39 & \textbf{0.40} & \textbf{0.33} \\
% \model & MAMMAL & CT & 500k &  --  & -- & -- & -- & -- & -- \\
% \model & MAMMAL & CT & 100k &  --  & -- & -- & -- & -- & -- \\
& & &  30k &  --  & 0.00 & 0.01 & 0.01 & 0.00 & 0.01 \\
\hline
% \model & ESM2-650M & AR & 10k &  --  & 0.02 & 0.05 & 0.08 & 0.11 & 0.07 \\
% \model & ESM2-150M & AR & 10k &  --  & 0.02 & 0.05 & 0.06 & 0.13 & 0.07 \\
% \model & ESM2-35M & AR  & 10k &  --  & 0.04 & 0.06 & 0.06 & 0.11 & 0.07 \\
% \model & ESM2-8M & AR   & 10k &  --  & 0.03 & 0.05 & 0.05 & 0.11 & 0.06 \\
\model & ESM2-8M & AR   & 100k &  --  & 0.21 & 0.12 & 0.20 & 0.24 & 0.19 \\
\hline
\smallerllm & \multicolumn{4}{c}{\multirow{4}{*}{(not applicable)}}  & 0.00 & 0.03 & 0.05 & 0.05 & 0.03 \\
Mixtral-8x7B & \multicolumn{4}{c}{}  & 0.00   & 0.02   & 0.06   & 0.02   & 0.02   \\
LLaMA-70B & \multicolumn{4}{c}{}  & 0.01   & 0.03   & 0.09   & 0.05   & 0.04   \\
GPT-OSS-120B & \multicolumn{4}{c}{}  & 0.03   & 0.09   & 0.06   & 0.10   & 0.07   \\
\bottomrule
\end{tabular}
\end{table}

\section{Discussion and Future Work}
\model demonstrates that \biofms and \llms can be aligned through lightweight projection layers, enabling generative reasoning across scRNA-seq, protein, and molecular modalities. This modular design allows compact LLMs to outperform much larger text-only baselines while producing richer, more interpretable outputs than candidate-matching approaches. By treating biological embeddings as first-class tokens, \model bridges raw data and language-based reasoning in a way that is both scalable and deployable.

A key strength of \model is its scalability across modalities: once aligned, scRNA-seq, proteins, and molecules can interoperate within the same \llm, supporting queries that span multiple levels of biology (e.g., “how does this variant protein affect cell type identity?” or “does this small molecule bind to this protein?”). Nonetheless, several limitations remain. The quality of alignment depends heavily on the underlying encoders, and modalities such as spatial transcriptomics or molecular 3D geometry are not yet explored. Current paired datasets rely largely on curated ontologies (e.g., GO terms, CellxGene metadata), which may bias reasoning and constrain coverage.

Looking ahead, several extensions are especially promising. First, interpretability can be enhanced by moving beyond single-token representations: gene-level, pathway-level, or topic-model embeddings (e.g., scETM \citep{scetm-zhao2021learning}, cisTopic \citep{bravo2019cistopic}) would yield more fine-grained rationales directly grounded in experimental data. Second, scaling to larger backbones (e.g., GPT-OSS-120B) and incorporating additional modalities such as epigenomics or spatial assays will test the limits of modularity and broaden biomedical applications. Third, standardized benchmarks are needed to evaluate not only accuracy but also interpretability, robustness, and factual grounding; multi-modal biological QA datasets remain scarce. Finally, integration into agentic workflows and privacy-preserving settings will be critical for real-world adoption. The design space is vast, and we have explored only a subset of configurations; further systematic ablations are essential. To accelerate progress, we will open-source our code and invite the community to co-develop multi-modal benchmarks and advance embedding-aware biomedical reasoning.

In summary, \model offers a unified and extensible framework for embedding-aware biomedical reasoning, laying the groundwork for practical systems that connect raw scientific data with natural language understanding and interactive discovery.

% ---------- Bibliography ----------
\bibliographystyle{iclr2026_conference}
\bibliography{references}

\begin{thebibliography}{34}
\providecommand{\natexlab}[1]{#1}
\providecommand{\url}[1]{\texttt{#1}}
\expandafter\ifx\csname urlstyle\endcsname\relax
  \providecommand{\doi}[1]{doi: #1}\else
  \providecommand{\doi}{doi: \begingroup \urlstyle{rm}\Url}\fi

\bibitem[Alayrac et~al.(2022)Alayrac, Donahue, Luc, Miech, Barr, Hasson, Lenc,
  Mensch, Millican, Reynolds, et~al.]{flamingo2022}
Jean-Baptiste Alayrac, Jeff Donahue, Pauline Luc, Antoine Miech, Iain Barr,
  Yana Hasson, Karel Lenc, Arthur Mensch, Katherine Millican, Malcolm Reynolds,
  et~al.
\newblock Flamingo: a visual language model for few-shot learning.
\newblock \emph{Advances in neural information processing systems},
  35:\penalty0 23716--23736, 2022.

\bibitem[Belcak et~al.(2025)Belcak, Heinrich, Diao, Fu, Dong, Muralidharan,
  Lin, and Molchanov]{belcak2025small}
Peter Belcak, Greg Heinrich, Shizhe Diao, Yonggan Fu, Xin Dong, Saurav
  Muralidharan, Yingyan~Celine Lin, and Pavlo Molchanov.
\newblock Small language models are the future of agentic ai.
\newblock \emph{arXiv preprint arXiv:2506.02153}, 2025.

\bibitem[Bian et~al.(2024)Bian, Chen, Dong, Li, Hao, Chen, Hu, Sun, Wei, and
  Zhang]{scmulan2024}
Haiyang Bian, Yixin Chen, Xiaomin Dong, Chen Li, Minsheng Hao, Sijie Chen,
  Jinyi Hu, Maosong Sun, Lei Wei, and Xuegong Zhang.
\newblock scmulan: a multitask generative pre-trained language model for
  single-cell analysis.
\newblock In \emph{International Conference on Research in Computational
  Molecular Biology}, pp.\  479--482. Springer, 2024.

\bibitem[Brandes et~al.(2022)Brandes, Ofer, Peleg, Rappoport, and
  Linial]{proteinbert2022}
Nadav Brandes, Dan Ofer, Yam Peleg, Nadav Rappoport, and Michal Linial.
\newblock Proteinbert: a universal deep-learning model of protein sequence and
  function.
\newblock \emph{Bioinformatics}, 38\penalty0 (8):\penalty0 2102--2110, 2022.

\bibitem[Bravo Gonz{\'a}lez-Blas et~al.(2019)Bravo Gonz{\'a}lez-Blas, Minnoye,
  Papasokrati, Aibar, Hulselmans, Christiaens, Davie, Wouters, and
  Aerts]{bravo2019cistopic}
Carmen Bravo Gonz{\'a}lez-Blas, Liesbeth Minnoye, Dafni Papasokrati, Sara
  Aibar, Gert Hulselmans, Valerie Christiaens, Kristofer Davie, Jasper Wouters,
  and Stein Aerts.
\newblock cistopic: cis-regulatory topic modeling on single-cell atac-seq data.
\newblock \emph{Nature methods}, 16\penalty0 (5):\penalty0 397--400, 2019.

\bibitem[Chen \& Zou(2024)Chen and Zou]{genept2024}
Yiqun Chen and James Zou.
\newblock Genept: a simple but effective foundation model for genes and cells
  built from chatgpt.
\newblock \emph{bioRxiv}, pp.\  2023--10, 2024.

\bibitem[Chithrananda et~al.(2020)Chithrananda, Grand, and
  Ramsundar]{chemberta2020}
Seyone Chithrananda, Gabriel Grand, and Bharath Ramsundar.
\newblock Chemberta: large-scale self-supervised pretraining for molecular
  property prediction.
\newblock \emph{arXiv preprint arXiv:2010.09885}, 2020.

\bibitem[Choi et~al.(2024)Choi, Park, Kim, Kim, Lee, Bae, Shin, and
  Lee]{cellama2024}
Hongyoon Choi, Jeongbin Park, Sumin Kim, Jiwon Kim, Dongjoo Lee, Sungwoo Bae,
  Haenara Shin, and Daeseung Lee.
\newblock Cellama: foundation model for single cell and spatial transcriptomics
  by cell embedding leveraging language model abilities.
\newblock \emph{bioRxiv}, pp.\  2024--05, 2024.

\bibitem[Cui et~al.(2024)Cui, Wang, Maan, Pang, Luo, Duan, and Wang]{scgpt2024}
Haotian Cui, Chloe Wang, Hassaan Maan, Kuan Pang, Fengning Luo, Nan Duan, and
  Bo~Wang.
\newblock scgpt: toward building a foundation model for single-cell multi-omics
  using generative ai.
\newblock \emph{Nature methods}, 21\penalty0 (8):\penalty0 1470--1480, 2024.

\bibitem[Dandala et~al.(2025)Dandala, Danziger, Barkan, Biswas, Gurev, Hu,
  Madgwick, Koseki, Kozlovski, Rosen-Zvi, et~al.]{bmfmrna2025}
Bharath Dandala, Michael~M Danziger, Ella Barkan, Tanwi Biswas, Viatcheslav
  Gurev, Jianying Hu, Matthew Madgwick, Akira Koseki, Tal Kozlovski, Michal
  Rosen-Zvi, et~al.
\newblock Bmfm-rna: An open framework for building and evaluating
  transcriptomic foundation models.
\newblock \emph{arXiv preprint arXiv:2506.14861}, 2025.

\bibitem[Fang et~al.(2023)Fang, Liang, Zhang, Liu, Huang, Chen, Fan, and
  Chen]{mol-instruct-fang2023}
Yin Fang, Xiaozhuan Liang, Ningyu Zhang, Kangwei Liu, Rui Huang, Zhuo Chen,
  Xiaohui Fan, and Huajun Chen.
\newblock Mol-instructions: A large-scale biomolecular instruction dataset for
  large language models.
\newblock \emph{arXiv preprint arXiv:2306.08018}, 2023.

\bibitem[Gayoso et~al.(2022)Gayoso, Lopez, Xing, Boyeau, Valiollah Pour~Amiri,
  Hong, Wu, Jayasuriya, Mehlman, Langevin, et~al.]{scvi-gayoso2022python}
Adam Gayoso, Romain Lopez, Galen Xing, Pierre Boyeau, Valeh Valiollah
  Pour~Amiri, Justin Hong, Katherine Wu, Michael Jayasuriya, Edouard Mehlman,
  Maxime Langevin, et~al.
\newblock A python library for probabilistic analysis of single-cell omics
  data.
\newblock \emph{Nature biotechnology}, 40\penalty0 (2):\penalty0 163--166,
  2022.

\bibitem[Jumper et~al.(2021)Jumper, Evans, Pritzel, Green, Figurnov,
  Ronneberger, Tunyasuvunakool, Bates, {\v{Z}}{\'\i}dek, Potapenko,
  et~al.]{alphafold2021}
John Jumper, Richard Evans, Alexander Pritzel, Tim Green, Michael Figurnov,
  Olaf Ronneberger, Kathryn Tunyasuvunakool, Russ Bates, Augustin
  {\v{Z}}{\'\i}dek, Anna Potapenko, et~al.
\newblock Highly accurate protein structure prediction with alphafold.
\newblock \emph{nature}, 596\penalty0 (7873):\penalty0 583--589, 2021.

\bibitem[Lee et~al.(2020)Lee, Yoon, Kim, Kim, Kim, So, and Kang]{biobert2020}
Jinhyuk Lee, Wonjin Yoon, Sungdong Kim, Donghyeon Kim, Sunkyu Kim, Chan~Ho So,
  and Jaewoo Kang.
\newblock Biobert: a pre-trained biomedical language representation model for
  biomedical text mining.
\newblock \emph{Bioinformatics}, 36\penalty0 (4):\penalty0 1234--1240, 2020.

\bibitem[Li et~al.(2023)Li, Li, Savarese, and Hoi]{blip2-2023}
Junnan Li, Dongxu Li, Silvio Savarese, and Steven Hoi.
\newblock Blip-2: Bootstrapping language-image pre-training with frozen image
  encoders and large language models.
\newblock In \emph{International conference on machine learning}, pp.\
  19730--19742. PMLR, 2023.

\bibitem[Lin et~al.(2023)Lin, Akin, Rao, Hie, Zhu, Lu, Smetanin, Verkuil,
  Kabeli, Shmueli, et~al.]{esm2_2023}
Zeming Lin, Halil Akin, Roshan Rao, Brian Hie, Zhongkai Zhu, Wenting Lu, Nikita
  Smetanin, Robert Verkuil, Ori Kabeli, Yaniv Shmueli, et~al.
\newblock Evolutionary-scale prediction of atomic-level protein structure with
  a language model.
\newblock \emph{Science}, 379\penalty0 (6637):\penalty0 1123--1130, 2023.

\bibitem[Liu et~al.(2023{\natexlab{a}})Liu, Li, Wu, and Lee]{llava2023}
Haotian Liu, Chunyuan Li, Qingyang Wu, and Yong~Jae Lee.
\newblock Visual instruction tuning.
\newblock \emph{Advances in neural information processing systems},
  36:\penalty0 34892--34916, 2023{\natexlab{a}}.

\bibitem[Liu et~al.(2023{\natexlab{b}})Liu, Li, Wang, Li, and
  Zhao]{sceval-liu2023}
Tianyu Liu, Kexing Li, Yuge Wang, Hongyu Li, and Hongyu Zhao.
\newblock Evaluating the utilities of foundation models in single-cell data
  analysis.
\newblock \emph{bioRxiv}, pp.\  2023--09, 2023{\natexlab{b}}.

\bibitem[Pei et~al.(2023)Pei, Zhang, Zhu, Wu, Gao, Wu, Xia, and Yan]{biot52023}
Qizhi Pei, Wei Zhang, Jinhua Zhu, Kehan Wu, Kaiyuan Gao, Lijun Wu, Yingce Xia,
  and Rui Yan.
\newblock Biot5: Enriching cross-modal integration in biology with chemical
  knowledge and natural language associations.
\newblock \emph{arXiv preprint arXiv:2310.07276}, 2023.

\bibitem[Perkel(2024)]{perkel2024cellxgene}
Jeffrey~M. Perkel.
\newblock 85 million cells — and counting — at your fingertips.
\newblock \emph{Nature}, 629:\penalty0 248--249, 2024.
\newblock \doi{10.1038/d41586-024-01217-y}.
\newblock URL \url{https://www.nature.com/articles/d41586-024-01217-y}.

\bibitem[Radford et~al.(2021)Radford, Kim, Hallacy, Ramesh, Goh, Agarwal,
  Sastry, Askell, Mishkin, Clark, et~al.]{clip2021}
Alec Radford, Jong~Wook Kim, Chris Hallacy, Aditya Ramesh, Gabriel Goh,
  Sandhini Agarwal, Girish Sastry, Amanda Askell, Pamela Mishkin, Jack Clark,
  et~al.
\newblock Learning transferable visual models from natural language
  supervision.
\newblock In \emph{International conference on machine learning}, pp.\
  8748--8763. PmLR, 2021.

\bibitem[Ross et~al.(2022)Ross, Belgodere, Chenthamarakshan, Padhi, Mroueh, and
  Das]{molformer2022}
Jerret Ross, Brian Belgodere, Vijil Chenthamarakshan, Inkit Padhi, Youssef
  Mroueh, and Payel Das.
\newblock Large-scale chemical language representations capture molecular
  structure and properties.
\newblock \emph{Nature Machine Intelligence}, 4\penalty0 (12):\penalty0
  1256--1264, 2022.

\bibitem[Schaefer et~al.(2024)Schaefer, Peneder, Malzl, Peycheva, Burton,
  Hakobyan, Sharma, Krausgruber, Menche, Tomazou, et~al.]{cellwhisperer2024}
Moritz Schaefer, Peter Peneder, Daniel Malzl, Mihaela Peycheva, Jake Burton,
  Anna Hakobyan, Varun Sharma, Thomas Krausgruber, Joerg Menche, Eleni~M
  Tomazou, et~al.
\newblock Multimodal learning of transcriptomes and text enables interactive
  single-cell rna-seq data exploration with natural-language chats.
\newblock \emph{bioRxiv}, pp.\  2024--10, 2024.

\bibitem[Shi et~al.(2025)Shi, Yang, Nai, Li, Fang, Wang, Liu, and
  Zhang]{scmmgpt-shi2025language}
Yaorui Shi, Jiaqi Yang, Changhao Nai, Sihang Li, Junfeng Fang, Xiang Wang,
  Zhiyuan Liu, and Yang Zhang.
\newblock Language-enhanced representation learning for single-cell
  transcriptomics.
\newblock \emph{arXiv preprint arXiv:2503.09427}, 2025.

\bibitem[Shoshan et~al.(2024)Shoshan, Raboh, Ozery-Flato, Ratner, Golts, Weber,
  Barkan, Rabinovici-Cohen, Polaczek, Amos, et~al.]{mammal2024}
Yoel Shoshan, Moshiko Raboh, Michal Ozery-Flato, Vadim Ratner, Alex Golts,
  Jeffrey~K Weber, Ella Barkan, Simona Rabinovici-Cohen, Sagi Polaczek, Ido
  Amos, et~al.
\newblock Mammal--molecular aligned multi-modal architecture and language.
\newblock \emph{arXiv preprint arXiv:2410.22367}, 2024.

\bibitem[Taylor et~al.(2022)Taylor, Kardas, Cucurull, Scialom, Hartshorn,
  Saravia, Poulton, Kerkez, and Stojnic]{galactica2022}
Ross Taylor, Marcin Kardas, Guillem Cucurull, Thomas Scialom, Anthony
  Hartshorn, Elvis Saravia, Andrew Poulton, Viktor Kerkez, and Robert Stojnic.
\newblock Galactica: A large language model for science.
\newblock \emph{arXiv preprint arXiv:2211.09085}, 2022.

\bibitem[Theodoris et~al.(2023)Theodoris, Xiao, Chopra, Chaffin, Al~Sayed,
  Hill, Mantineo, Brydon, Zeng, Liu, et~al.]{geneformer2023}
Christina~V Theodoris, Ling Xiao, Anant Chopra, Mark~D Chaffin, Zeina~R
  Al~Sayed, Matthew~C Hill, Helene Mantineo, Elizabeth~M Brydon, Zexian Zeng,
  X~Shirley Liu, et~al.
\newblock Transfer learning enables predictions in network biology.
\newblock \emph{Nature}, 618\penalty0 (7965):\penalty0 616--624, 2023.

\bibitem[Wang et~al.(2025{\natexlab{a}})Wang, Schmidgall, Jaeger, Zhang,
  Pilgrim, Matias, Barral, Fleet, and Azizi]{txgemma2025}
Eric Wang, Samuel Schmidgall, Paul~F Jaeger, Fan Zhang, Rory Pilgrim, Yossi
  Matias, Joelle Barral, David Fleet, and Shekoofeh Azizi.
\newblock Txgemma: Efficient and agentic llms for therapeutics.
\newblock \emph{arXiv preprint arXiv:2504.06196}, 2025{\natexlab{a}}.

\bibitem[Wang et~al.(2025{\natexlab{b}})Wang, Gao, Gu, Pu, Cui, Wei, Liu, Jing,
  Ye, Shao, et~al.]{internvl3-2025}
Weiyun Wang, Zhangwei Gao, Lixin Gu, Hengjun Pu, Long Cui, Xingguang Wei,
  Zhaoyang Liu, Linglin Jing, Shenglong Ye, Jie Shao, et~al.
\newblock Internvl3. 5: Advancing open-source multimodal models in versatility,
  reasoning, and efficiency.
\newblock \emph{arXiv preprint arXiv:2508.18265}, 2025{\natexlab{b}}.

\bibitem[Yang et~al.(2022)Yang, Wang, Wang, Fang, Tang, Huang, Lu, and
  Yao]{scbert2022}
Fan Yang, Wenchuan Wang, Fang Wang, Yuan Fang, Duyu Tang, Junzhou Huang, Hui
  Lu, and Jianhua Yao.
\newblock scbert as a large-scale pretrained deep language model for cell type
  annotation of single-cell rna-seq data.
\newblock \emph{Nature Machine Intelligence}, 4\penalty0 (10):\penalty0
  852--866, 2022.

\bibitem[Yu et~al.(2024)Yu, Baker, Chen, Ning, and Sun]{llasmol-yu2024}
Botao Yu, Frazier~N Baker, Ziqi Chen, Xia Ning, and Huan Sun.
\newblock Llasmol: Advancing large language models for chemistry with a
  large-scale, comprehensive, high-quality instruction tuning dataset.
\newblock \emph{arXiv preprint arXiv:2402.09391}, 2024.

\bibitem[Yuan et~al.(2024)Yuan, Zhan, Zhang, Zhou, Zhao, Han, Li, and
  Tang]{sccello2024}
Xinyu Yuan, Zhihao Zhan, Zuobai Zhang, Manqi Zhou, Jianan Zhao, Boyu Han, Yue
  Li, and Jian Tang.
\newblock Cell ontology guided transcriptome foundation model.
\newblock \emph{Advances in Neural Information Processing Systems},
  37:\penalty0 6323--6366, 2024.

\bibitem[Zhao et~al.(2024)Zhao, Zhang, Wu, Luo, and Nie]{zhao2024-langcell}
Suyuan Zhao, Jiahuan Zhang, Yushuai Wu, Yizhen Luo, and Zaiqing Nie.
\newblock Langcell: Language-cell pre-training for cell identity understanding.
\newblock \emph{arXiv preprint arXiv:2405.06708}, 2024.

\bibitem[Zhao et~al.(2021)Zhao, Cai, Zhang, Tang, and
  Li]{scetm-zhao2021learning}
Yifan Zhao, Huiyu Cai, Zuobai Zhang, Jian Tang, and Yue Li.
\newblock Learning interpretable cellular and gene signature embeddings from
  single-cell transcriptomic data.
\newblock \emph{Nature communications}, 12\penalty0 (1):\penalty0 5261, 2021.

\end{thebibliography}

\end{document}